\documentclass[aps,prd,twocolumn,showpacs,superscriptaddress,letterpaper,floatfix,pdftex]{revtex4}
\usepackage{graphicx}
\usepackage{epsfig}

\usepackage{longtable}
\usepackage{amssymb}
\usepackage{amsthm}
\usepackage{amsfonts}
\usepackage[centertags]{amsmath}
\usepackage{color}
\usepackage{sidecap}
\usepackage{pst-all}
\usepackage{pstricks}
\usepackage[latin1]{inputenc}
\usepackage[english]{babel}
\usepackage{soul}
\usepackage{pifont} 

\usepackage{array}
\usepackage{xspace}
\usepackage{ulem} 
\usepackage{verbatim}

\newcommand{\I}{\ensuremath{{\rm i}\,}}

\newcommand{\LG}{Laguerre Gauss\xspace}

\newcommand{\IFO}{interferometer\xspace}

\newcommand{\ADV}{Advanced Virgo\xspace}

\begin{document}

\title[Prospects of higher-order \LG modes in future GW detectors]{Prospects of higher-order \LG modes in future gravitational wave detectors}

\author{Simon Chelkowski}
\affiliation{School of Physics and Astronomy, University of
Birmingham, Edgbaston, Birmingham B15 2TT, UK}
\author{Stefan Hild}
\affiliation{School of Physics and Astronomy, University of
Birmingham, Edgbaston, Birmingham B15 2TT, UK}
\author{Andreas Freise}
\affiliation{School of Physics and Astronomy, University of
Birmingham, Edgbaston, Birmingham B15 2TT, UK}

\date{\today}

\begin{abstract}
The application of higher-order Laguerre Gauss (LG) modes in
large-scale gravitational wave detectors has recently been
proposed. In comparison to the fundamental mode, some higher-order
Laguerre Gauss modes can significantly reduce the contribution of
coating Brownian noise. 
Using frequency domain simulations we give a detailed analysis of
the longitudinal and angular control signals derived with a
LG$_{33}$ mode in comparison to the fundamental TEM$_{00}$ mode.
The performance regarding interferometric sensing and control of
the LG$_{33}$ mode is found to be similar, if not even better in
all aspects of interest. 
In addition, we evaluate the sensitivity gain of the
implementation of LG$_{33}$ modes into the Advanced Virgo
instrument. Our analysis shows that the application of the
LG$_{33}$ mode results in a broadband improvement of the Advanced
Virgo sensitivity, increasing the potential detection rate of
binary neutron star inspirals by a factor 2.1.
\end{abstract}

\pacs{04.80.Nn, 95.75.Kk, 42.25.Bs, 95.55.Ym}


\maketitle

\section{Introduction}

The search for gravitational waves (GW) has led to a new class of
extremely sensitive laser interferometers.  The first generation
of large-scale laser-interferometric gravitational wave detectors
\cite{geo2006, virgo2006short, ligo2004, tama2004} is now in
operation with the aim of accomplishing the first direct detection
of gravitational waves. The detector performance is limited by
several fundamental and technical noises. In a constant effort the
noise contributions are minimised to improve the detectors
signal-to-noise ratio. One of the limiting noise sources of the
currently planned second generation gravitational wave detectors
will be thermal noise \cite{VIR-101A-08} of the mirror test
masses. There exist several components to thermal noise of which
the Brownian thermal noise is largest in current interferometer
topologies utilising arm cavities. Cooling of the mirror test
masses as currently studied in CLIO \cite{CLIO08} reduces the
thermal noise provided an appropriate material is chosen for the
optics. A different way to lower the thermal noise is to change
the mode shape of the laser beam inside the interferometer. All
current detectors use the fundamental TEM$_{00}$ mode, but several
other mode shapes such as Mesa beams \cite{BondarescuThorne06},
conical modes \cite{Bondarescu08} and higher-order Laguerre Gauss
(LG) modes \cite{Mours06} have been proposed for reducing thermal
noise. The basic idea is to reduce thermal noise by generating a
more uniform light intensity distribution on the mirrors without
introducing higher clipping losses \cite{Mours06}.

The proposed candidates for different beam shapes can be divided
in two groups. The first group, which comprises flat-top and
conical beams, would require the use of non spherical mirror
shapes. As a result these modes are not compatible with current GW
detectors and their spherical mirrors. Currently it is not clear
to what precision these non-spherical mirrors can be manufactured
and little experience in using such mirrors has been gained so far. 
The second group consists of higher-order Laguerre Gauss modes,
which are fully compatible with spherical mirrors as the 
currently used TEM$_{00}$ mode\footnote{Please note that we use
the equivalent terms \textit{fundamental mode}, \textit{TEM$_{00}$
mode}, \textit{HG$_{00}$ mode} and \textit{LG$_{00}$ mode}
throughout this paper, depending on which of these seems to be
more appropriate in the context.}. So far LG modes have been
mainly employed in the field of cold atom and quantum optics for
example as optical tweezers \cite{HFHR95} or waveguides
\cite{BBDHAES01}.

Currently several techniques for the generation of higher-order
LG-modes exist e.g. using holograms \cite{Arlt98,Clifford98},
gratings \cite{Kennedy02} and mode transformers
\cite{Courtial99,ONeil00}. With these techniques a conversion
efficiency of 60\% \cite{Kennedy02} has been demonstrated.
Recently the creation of higher-order LG-modes with a very high
mode purity \cite{Chu08} has been reported. Our paper assumes that
using these techniques, higher-order LG-modes can be created with
high power output and high mode purity required in the field of GW
detection. We analyse the compatibility of such higher-order
LG-modes with the core interferometer in future GW detectors,
using \ADV in particular as an example for a second-generation GW
detector. In Section~\ref{sec:initial-considerations} we introduce
the definition of the LG-modes, how they can be described in a
Hermite Gauss mode basis system and how the coating brownian
thermal noise is calculated depending on which LG-mode is used.
Moreover we introduce some practical considerations concerning
clipping loss, beam sizes and radius of curvature (RoC) of the
mirrors which are essential for the later analysis. In
Section~\ref{sec:phase-coupling-analysis} we perform a phase
coupling analysis of a single arm cavity and a Michelson
interferometer using higher-order LG modes in comparison to the
currently used fundamental mode. We determine any differences in
their phase coupling between the different longitudinal and
alignment degrees of freedom. In
Section~\ref{sec:prospects-Adv-Virgo} a numerical model based on a
set of \ADV design parameters is used to analyse the prospects of
higher-order LG modes in comparison to the currently proposed use
of the fundamental mode. The detector sensitivities of the
different \IFO configurations are computed to derive the
envisioned detector inspiral ranges. In total we analyse and
compare three different cases with each other.

\section{Initial considerations}
\label{sec:initial-considerations}

\subsection{Hermite Gauss and Laguerre Gauss modes}
The Hermite Gauss (HG) modes and Laguerre Gauss modes both present
complete basis sets such that each LG mode can be presented by a
sum of HG modes and vice versa. The so-called helical Laguerre
Gauss modes can be written as \cite{Siegman,SiegmanErrata}:
\begin{eqnarray}\label{eq:uld1}
u_{p,l}&(r,\phi,z)= \frac{1}{w(z)}\sqrt{\frac{2p!}{\pi(|l|+p)!}}\exp(\I(2p+|l|+1)\Psi(z))\nonumber\\
&\times\left(\frac{\sqrt{2}r}{w(z)}\right)^{|l|}L_p^l\left(\frac{2r^2}{w(z)^2}\right)
\exp\left(-\I k\frac{r^2}{2q(z)}+\I l \phi\right)
\end{eqnarray}
with $r$, $\phi$ and $z$ as the cylindrical coordinates around the
optical axis, $w(z)$ the beam radius, $\Psi(z)$ the Gouy phase,
$q(z)$ the Gaussian beam parameter and $L_p^l(x)$ the associated
Laguerre polynomials. The indices must obey the following
relations: $0\leq |l|\leq p$ where $p$ is the radial mode index
and $l$ the azimuthal mode index.

The decomposition of these modes into Hermite Gauss modes can be
performed as follows \cite{Beijersbergen93}:
\begin{equation}\label{eq:HGLGconversion}
u_{p,l}(x,y,z)=\sum_{k=0}^{(2p+1)} \I^k b(l+p,p,k)
u^{HG}_{2p+l-k,k}(x,y,z)
\end{equation}
with real coefficients
\begin{eqnarray}
b(l+p,p,k)&=&\sqrt{\frac{(2p+l-k)!k!}{2^{(2p+l)}
(l+p)!p!}}\nonumber\\&\phantom{=}&\quad\times(-2)^k
P_k^{l+p-k,p-k}(0)
\end{eqnarray}
where $P_n^{\alpha, \beta}(x)$ denotes the Jacobi Polynomials. It
is interesting to note in Equation~\ref{eq:HGLGconversion} that a
given LG$_{pl}$ mode is constructed of $2p+l+1$ HG$_{nm}$ modes of
the order $n+m=2p+l$. For example the LG$_{33}$ mode is
constructed out of ten Hermite Gauss modes of the order nine.

\subsection{Coating brownian thermal noise of Laguerre-Gauss modes}
According to \cite{Levin98} the power spectral density of
displacement equivalent thermal noise is generally given by
\begin{equation}\label{spectral-density-of-displacement}
S_x(f)=\frac{4\,k_B\,T}{\pi\,f}\,\phi\,U\,,
\end{equation}
with $\phi$ being the loss angle and U the strain energy of the
static pressure profile on the mirror surface normalised to 1\,N.
The interested reader is referred to \cite{Vinet09} where detailed
calculations of various thermal noises such as substrate brownian,
coating brownian and thermoelastic thermal noise are presented.
The currently most limiting thermal noise in GW detectors is the
coating brownian thermal noise. In the case of a semi-infinite
mirror the coating brownian thermal noise induced by a LG$_{pl}$
mode can be calculated using the strain energy
\begin{equation}\label{U-coating}
U_{p,l,\mathrm{coating}}=\delta_c\frac{(1+\sigma)(1-2\sigma)}{\pi\,Y\,w^2}\,g_{p,l}\,.
\end{equation}
Here $\delta_c$ is the thickness of the coating, $\sigma$ is the
Poisson ratio, $Y$ is the Young modulus, $w$ is the beam width at
the mirror and $g_{p,l}$ is a scaling factor depending on the used
LG$_{pl}$ mode. In the case of the fundamental LG$_{00}$ mode this
scaling factor is $g_{0,0}=1$, whereas for a LG$_{33}$ mode
$g_{3,3}=0.14$ has to be used. Hence the power spectral density of
displacement equivalent coating brownian thermal noise is more
than a factor of seven smaller for a LG$_{33}$ mode in comparison
to the fundamental LG$_{00}$ mode. For finite mirror sizes
\cite{Vinet09} finds that the deviation to the semi-infinite case
are very small if the clipping loss of the beam on the mirror
surface are small.

\subsection{Clipping loss and beam scaling factors}\label{sec:clipping-loss}

For our later analysis it is important to know the clipping loss
$l_\mathrm{clip}$ that affects the propagating Gaussian mode at
the mirror due to its finite size. It is given by
\begin{eqnarray}
  l_\mathrm{clip}(w,\rho,z)&=&1-\int_0^{2\pi}d\phi\,\int_0^\rho dr\cdot r \nonumber\\
  &\phantom{=}&\quad\times \,u(w,r,\phi,z)\, u^*(w,r,\phi,z)\,,
\label{clipping-loss}
\end{eqnarray}
were $w$ is the beam radius at the mirror, $\rho$ is the radius of
the mirror coating and $u(w,r,\phi,z)$ is the transversal field
distribution of the mode of interest. Please note that the
parameter beam radius $w$ is a measure of the beam size of the
fundamental Gaussian mode (LG$_{00}$ or HG$_{00}$). Higher-order
LG or HG modes of the same beam radius actually are more spatially
extended, in the sense that a significant amount light power can
be detected at distances off the optical axis larger than the beam
radius. Using the general definitions of the transversal field
distribution for Hermite Gauss and Laguerre Gauss modes $u$ we can
compute the clipping losses of any of these modes.

\begin{figure}[htb]
\centerline{\includegraphics[width=8.6cm,keepaspectratio]{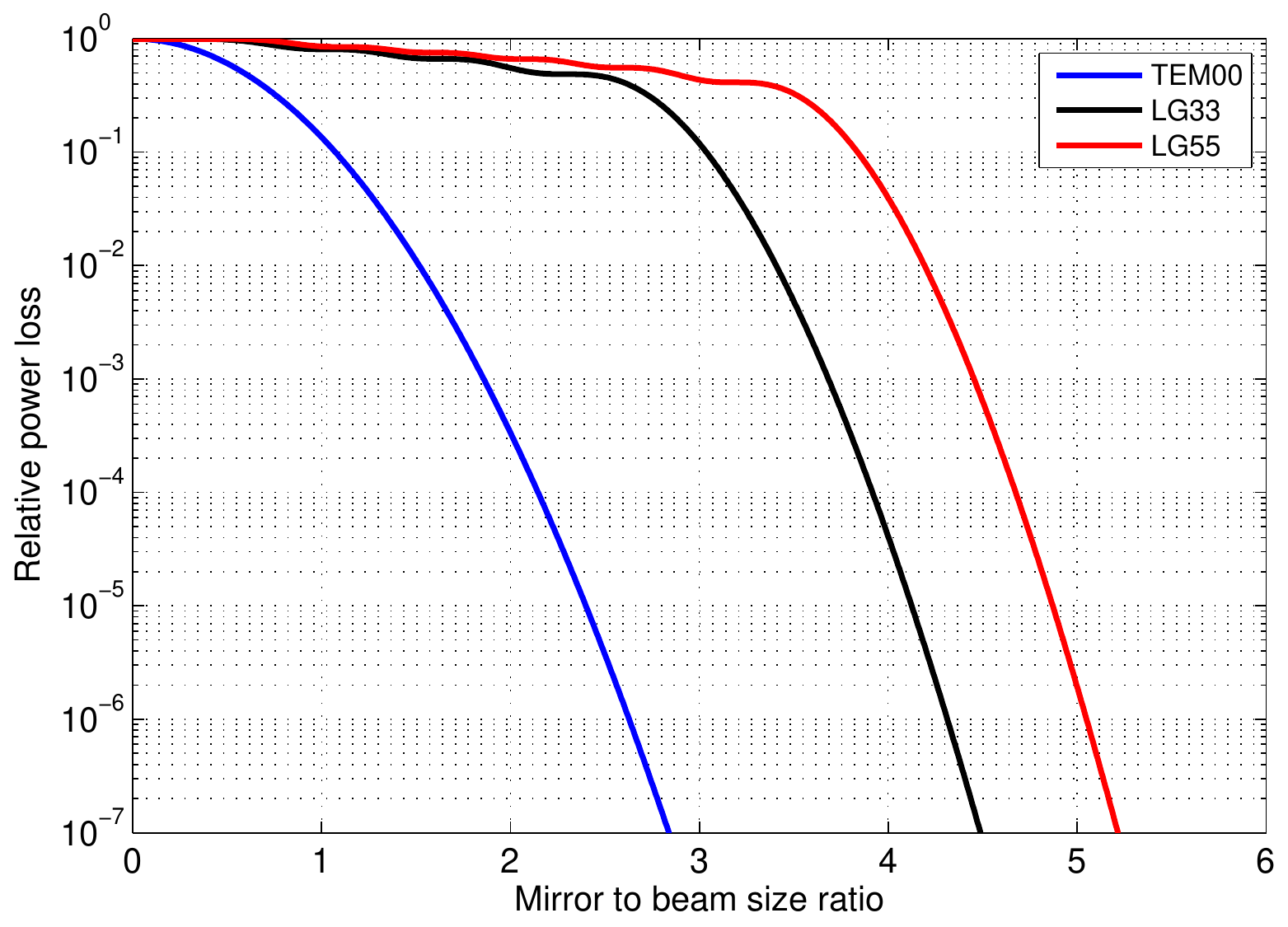}}
\caption{The plot shows the relative power loss in reflection of a
finite sized mirror due to clipping loss for three different
incident transversal modes over the mirror to beam radius ratio.
The different curves shown are: blue curve = TEM$_{00}$ mode;
black curve = LG$_{33}$ mode; red curve = LG$_{55}$ mode. The
mirror to beam radius ratio for a fixed clipping loss of 1\,ppm is
given in Table~\ref{tab:beam-scaling-factors}}
\label{relative_power_loss_same_radius.eps}
\end{figure}

In Figure~\ref{relative_power_loss_same_radius.eps} the clipping
losses for the fundamental and two higher-order LG modes are
plotted over the mirror-radius to beam-radius ratio. One can see
clearly that in comparison to the fundamental LG$_{00}$ mode, the
higher-order LG modes have a much more widely spread intensity
distribution for a given beam radius. Hence they require either
larger mirrors, or reduced beam radii for a fixed mirror size.
Table~\ref{tab:beam-scaling-factors} comprises the respective
scaling factors, normalised to a optimised mirror size for a
LG$_{00}$--mode with a clipping loss of 1\,ppm.
\begin{table}[!h]
\begin{center}
\begin{tabular}{l|c|c|c}
&  LG$_{00}$ &  LG$_{33}$ &  LG$_{55}$\\
\hline\hline
mirror to beam radius ratio 
 & 2.63 & 4.31 & 5.05 \\
relative mirror radius & 1 & 1.64 & 1.92 \\
relative beam radius   & 1 & 0.61 & 0.52 \\
\end{tabular}
\end{center}
\caption{Comparison of the mirror to beam radius ratio for 1\,ppm
clipping loss and the corresponding scaling factors for the beam
radius and mirror radius to keep the clipping loss constant
normalised to the LG$_{00}$--mode with a clipping loss of 1\,ppm.}
\label{tab:beam-scaling-factors}
\end{table}

\noindent One consequence of this more widely spread intensity
distribution of the higher-order LG modes is of major importance
for the later analysis: in order to fit a higher-order mode
optimally on the same mirror as the fundamental mode, the beam
radius of the higher-order mode must be different from that of the
fundamental mode. This corresponds to a different wave front
curvature and consequently to a different spherical curvature of
the cavity mirrors \cite{Siegman}. Thus changing an existing
optical experiment such as an interferometer from a configuration
using e.g. the TEM$_{00}$ mode to a configuration using the
LG$_{33}$ mode, the radii of curvature of the mirrors must be
changed, if one wants to keep the clipping losses at a constant
level. In most cases this would necessitate in a complete exchange
of the mirrors.

\section{Phase coupling comparison of the LG$_{33}$ mode with the fundamental HG$_{00}$ mode }\label{sec:phase-coupling-analysis}

The sensitivity of future gravitational wave detectors will be
limited partly by thermal noise. The use of higher-order LG modes
represents a very interesting option for reducing this limit. For
a successful implementation, however, higher-order LG modes must
comply with the stringent phase noise requirements in these
detectors. In the following we compare the phase noise performance
of the currently used HG$_{00}$ mode with that of a LG$_{33}$ mode
which serves as a representative of the family of higher-order LG
modes. The analysis was performed with the numerical
interferometer simulation \textsc{Finesse} \cite{Finesse}, which
uses the Hermite Gauss modal expansion for describing the spatial
properties of light fields transverse to the optical axis. In
order to simulate higher-order LG modes we used the decomposition
of higher-order LG into HG modes presented in
Section~\ref{sec:initial-considerations}. The results of the
analysis described in this section are in principle applicable to
many other Laguerre Gauss modes of interest. For instance for the
LG$_{55}$ mode, the corresponding mirror and beam radius scaling
values are given in table~\ref{tab:beam-scaling-factors}.

\subsection{Configurations of interest}\label{sec:configurations-of-interest}

\begin{figure*}[t]
\centerline{
\includegraphics[width=17.8cm,keepaspectratio]{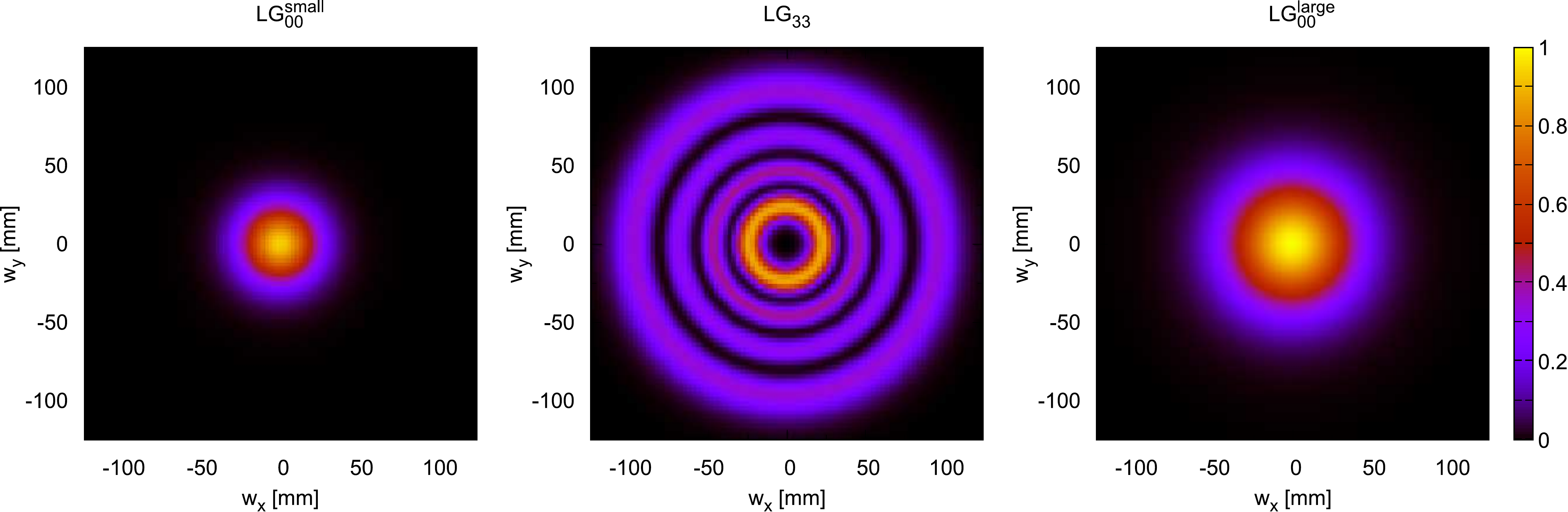}
} \caption{Transversal intensity distribution at the mirror
surface for the three configurations under investigation.}
\label{modeshapes.eps}
\end{figure*}

Our phase noise coupling analysis uses the currently planned 3\,km
long \ADV interferometer as testbed. We compare the use of a
LG$_{33}$ mode with two different configurations using a
fundamental mode:
\begin{enumerate}
\renewcommand{\labelenumi}{\textbf{\Roman{enumi}.}}
\renewcommand{\labelenumi}{\textbf{\arabic{enumi}.)}}
\item The $\mathrm{LG_{00}^{small}}$ configuration: This
configuration uses the optical parameters presented in
\cite{VIR-NOT-EGO-1390-330} and represents our reference
configuration. The configuration uses arm cavity mirrors with RoCs
$R_C=\pm 1910$\,m which corresponds to a beam size of the
fundamental $\mathrm{LG_{00}}$ mode of $w=35.2$\,mm at the mirrors
surface and a corresponding waist size of $w_0=16.3$\,mm.
According to table~\ref{tab:beam-scaling-factors} this
configuration would have a clipping loss of 1\,ppm for a mirror
radius of $\rho_\mathrm{small}=92.5$\,mm.

\item The $\mathrm{LG_{33}}$ configuration: This configuration
uses the higher-order Laguerre Gauss mode $\mathrm{LG_{33}}$. It
shares its beam parameters with the reference
$\mathrm{LG_{00}^{small}}$ configuration to simplify the
comparison. The mirror radius has to be adapted for this
configuration because of the more wider intensity profile of the
$\mathrm{LG_{33}}$ mode.
A mirror radius of $\rho_\mathrm{large}=151.8$\,mm is required to
maintain a clipping loss of 1\,ppm (see
Table~\ref{tab:beam-scaling-factors}).

\item The $\mathrm{LG_{00}^{large}}$ configuration: The third
configuration uses the fundamental $\mathrm{LG_{00}}$ mode in
combination with the larger mirrors with radius
$\rho_\mathrm{large}$ used in the $\mathrm{LG_{33}}$
configuration. As a result the beam size on the mirrors can be
increased to $w=57.7$\,mm while still maintaining a clipping loss
of 1\,ppm. Hence the waist size decreases to $w_0=8.9$\,mm and all
other beam parameters change accordingly.
\end{enumerate}
The beam parameters of each configuration are displayed in
Table~\ref{tab:configurations_of_interest} and their transversal
intensity distribution on the mirror surface is shown in
Figure~\ref{modeshapes.eps}. It is worth noting, that a comparison
between the $\mathrm{LG_{33}}$ and the $\mathrm{LG_{00}^{large}}$
configuration is much more reasonable because these two
configurations use the same mirror sizes.

\begin{table}[!h]
\begin{center}
\begin{tabular}{l|c|c|c}
   & $\mathrm{LG_{00}^{small}}$ & $\mathrm{LG_{33}}$ & $\mathrm{LG_{00}^{large}}$\\
  \hline
  \hline
  $R_C$ [m] & 1910 & 1910 & 1536.7 \\
  $w$ [mm] & 35.2 & 35.2 & 57.7 \\
  $w_0$ [mm] & 16.3 & 16.3 & 8.9 \\
\end{tabular}
\end{center}
\caption{Beam and mirror parameters of the three different
configurations used in the analysis. There exist always two RoC
settings for achieving a given spot size on the mirrors. We have
chosen the cavity geometry which reduces the radiation pressure
induced alignment instabilities~\cite{sidles06}.}
\label{tab:configurations_of_interest}
\end{table}


\subsection{Tilt to longitudinal phase coupling of a single cavity}
\label{sec:sim-single-cavity}

The first part of our analysis is performed for a single cavity.
At first we want to investigate the longitudinal error signal of
each configuration in order to find out how they compare against
each other. To generate this longitudinal error signal we use the
Pound-Drever-Hall technique based on a modulation/demodulation
scheme \cite{Black00}. We found that all resulting error signals
are identical, which confirms that such error signals only depend
on the average phase of the beam, and is independent of its modal
distribution.

Next we analyse the coupling of misalignment of a cavity into
longitudinal phase noise for the three configurations of interest.
\begin{figure*}[t]
\centerline{
\includegraphics[width=17.8cm,keepaspectratio]{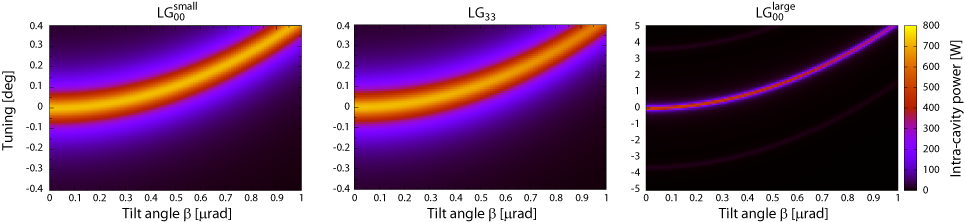}}
\caption{Intra cavity power over tilt angle $\beta$ of the end
mirror (EMX) and longitudinal tuning $\phi$ of a single cavity
shown for all three configurations of interest. The first two
configurations show almost the same coupling from tilt into
longitudinal tuning which is more than an order of magnitude lower
than the coupling of the third configuration.}
\label{2D-coupling-tilt-into-longitudinal.eps}
\end{figure*}
The first results are comprised in
Figure~\ref{2D-coupling-tilt-into-longitudinal.eps} which shows
the intra cavity power over the tilt angle $\beta$ of the end
mirror (EMX) and the longitudinal tuning $\phi$ of the cavity for
each configuration. The tuning value $\phi$ of the cavity is the
result of a modulo division of the cavity length by the
wavelength. In the following this tuning is given in degrees with
$360^\circ$ referring to one wavelength:
\begin{equation}
  \phi=360^\circ \cdot \left(L\quad\mathrm{mod}\quad\lambda\right)
\label{cavity_tuning}
\end{equation}
In the analysis the tilt angle $\beta$ of the cavity's end mirror
--- shown on the x-axis
--- was varied between 0\,$\mu$rad to 1\,$\mu$rad. On the y-axis
the tuning $\phi$ was chosen such that the resonance of the cavity
is clearly visible. One can see that a tilt of the end mirror
changes the tuning of the resonance for every configuration.
Because the tuning refers to the length of the cavity, there is
indeed a coupling from tilt into the longitudinal phase degree of
freedom of the cavity. The coupling strength is different for each
configuration. Nevertheless, configuration
$\mathrm{LG_{00}^{small}}$ and $\mathrm{LG_{33}}$ behave in a very
similar manner. Both show a shift of the resonance condition of
approximately $\Delta\phi=0.4^\circ$ for a tilt of 1\,$\mu$rad. In
contrast, configuration $\mathrm{LG_{00}^{large}}$ shows an
increased coupling strength by more than one order of magnitude.
For a tilt of 1\,$\mu$rad the cavity tuning of for the
$\mathrm{LG_{00}^{large}}$  configuration changes by
$\Delta\phi\approx 5^\circ$.

\begin{figure}[t]
\centerline{
\includegraphics[width=8.6cm,keepaspectratio]{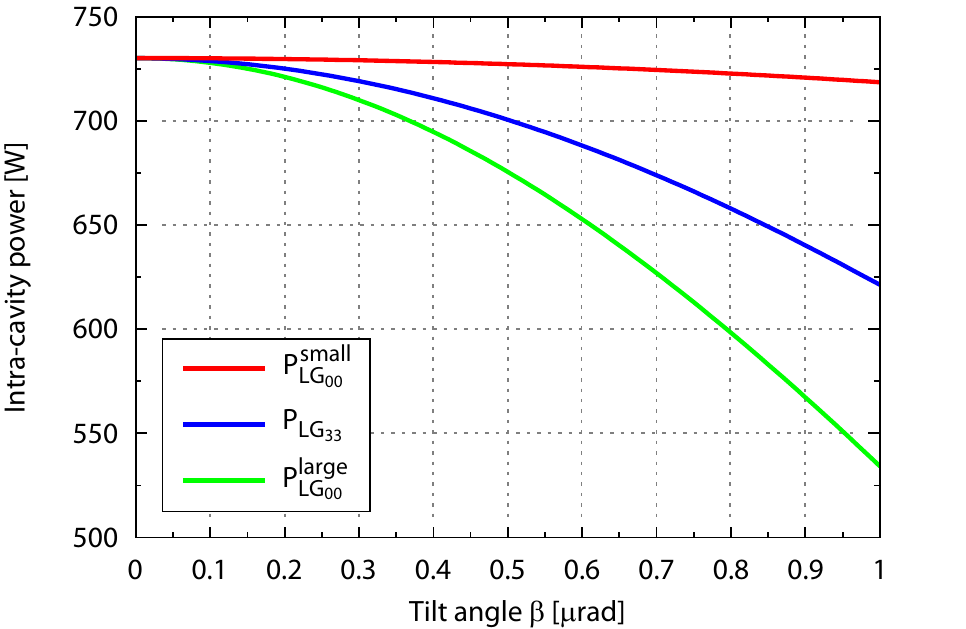}
} \caption{Comparison between all three cases and their coupling
between tilt of mirror EMX to the tuning of mirror IMX.}
\label{tilt-into-intra-cavity-power-coupling.eps}
\end{figure}
The tilt of the end mirror (EMX) also changes the geometry of the
eigenmode of the cavity. Compared to the input beam the eigenmode
of the cavity is tilted as well. As a result the mode matching
efficiency into the cavity is decreased, leading to a reduced
intra cavity power. This behaviour is hardly visible in
Figure~\ref{2D-coupling-tilt-into-longitudinal.eps} but clearly
shown in Figure~\ref{tilt-into-intra-cavity-power-coupling.eps}.
The $\mathrm{LG_{00}^{small}}$ configuration is the most robust in
terms of decreased intra cavity power, followed by the
$\mathrm{LG_{33}}$ and the $\mathrm{LG_{00}^{large}}$
configurations.

In conclusion of this section the $\mathrm{LG_{00}^{small}}$
configuration performs best in all aspects of the analysis.
Nevertheless as stated initially a comparison between the other
two configurations is much fairer because they share the same
mirror size. Taking this into account, the favourable mode is the
$\mathrm{LG_{33}}$, because its tilt induced coupling into the
longitudinal phase and into the intra cavity power is much less
than for a $\mathrm{LG_{00}}$ mode on the same mirrors.

\subsection{Alignment analysis of a single arm
cavity}\label{sec:alignment_analysis}

\begin{figure*}[t]
\centerline{
\includegraphics[width=16cm,keepaspectratio]{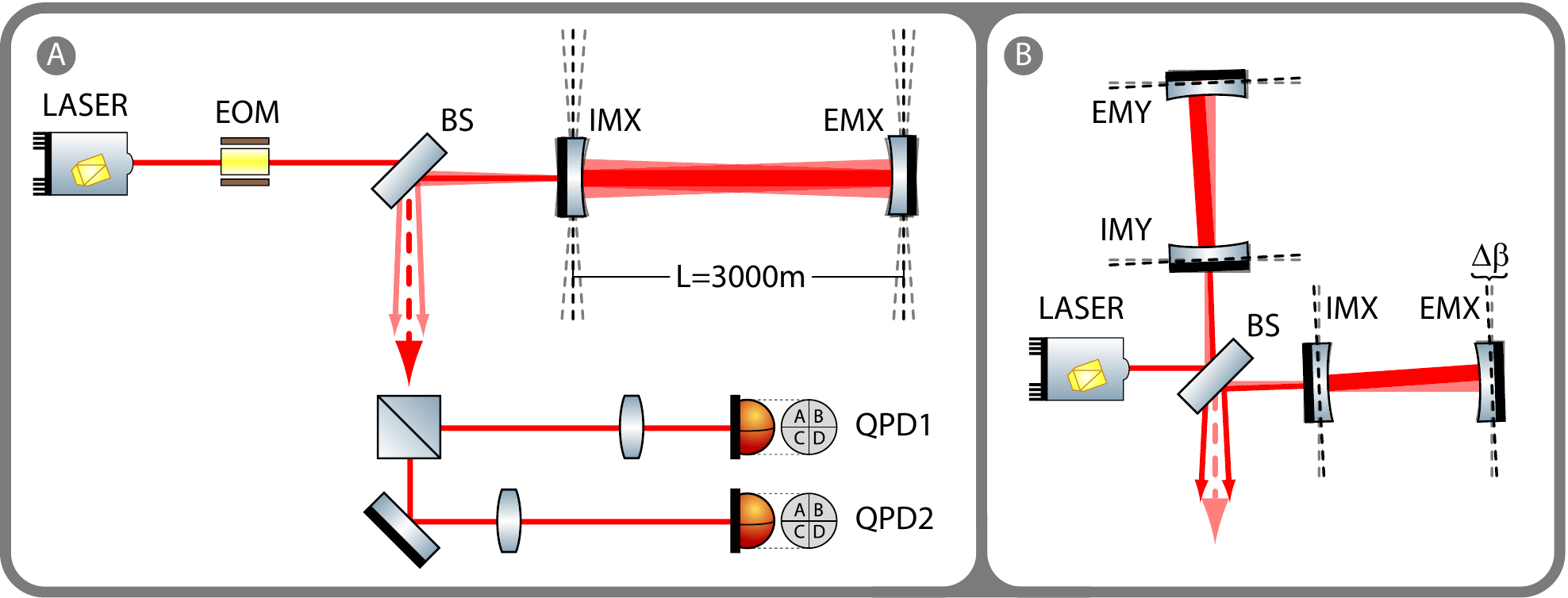}
} \caption{Different optical layouts used in the alignment
analysis described in Section~\ref{sec:alignment_analysis}. Left
\textcircled{\scriptsize{\textbf{A}}}: The generation of alignment
error signals for a single arm cavity. Right
\textcircled{\scriptsize{\textbf{B}}}: Michelson interferometer
with differentially misaligned arm cavities to study the power
coupling into the output port of the interferometer.}
\label{paper-cavity-alignment.eps}
\end{figure*}

\begin{figure}[t]
\includegraphics[width=8.6cm,keepaspectratio]{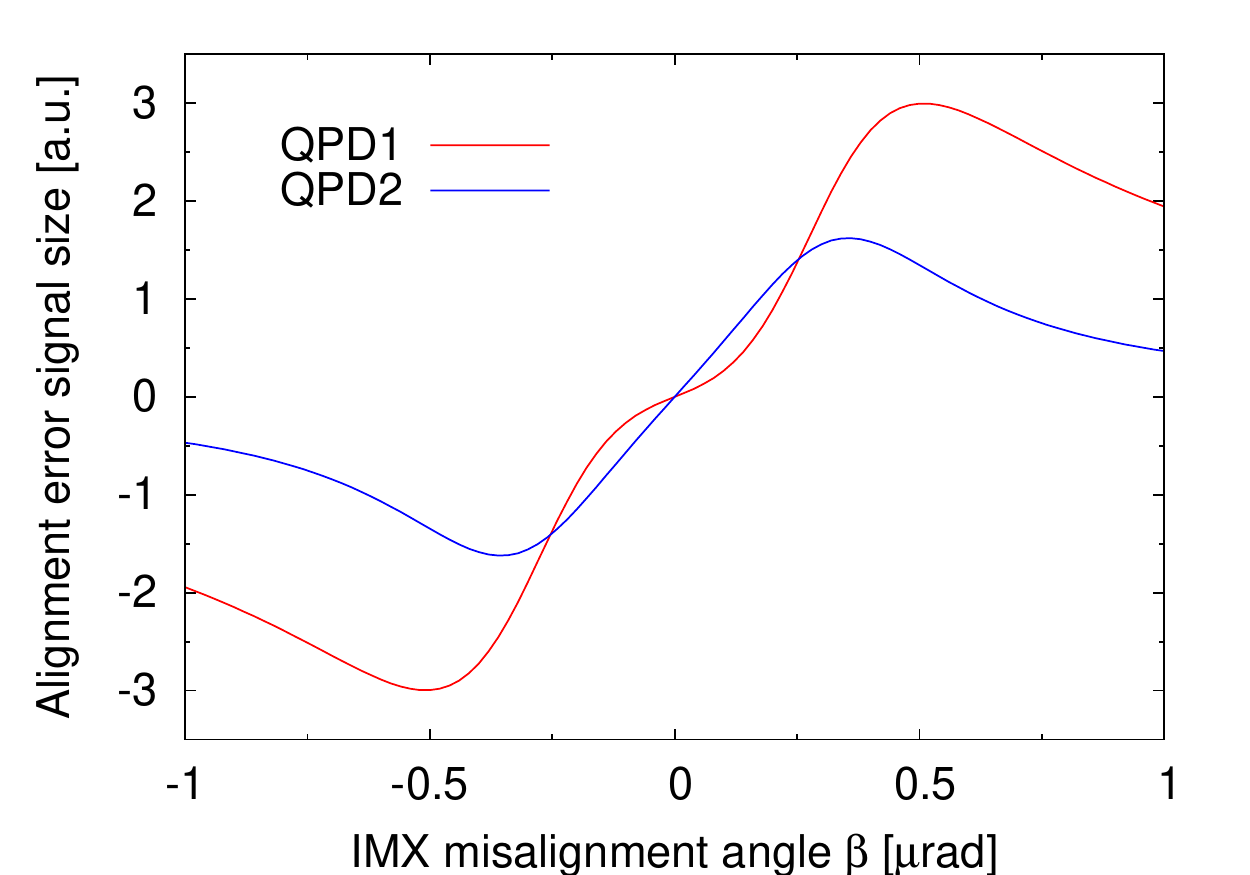}
\includegraphics[width=8.6cm,keepaspectratio]{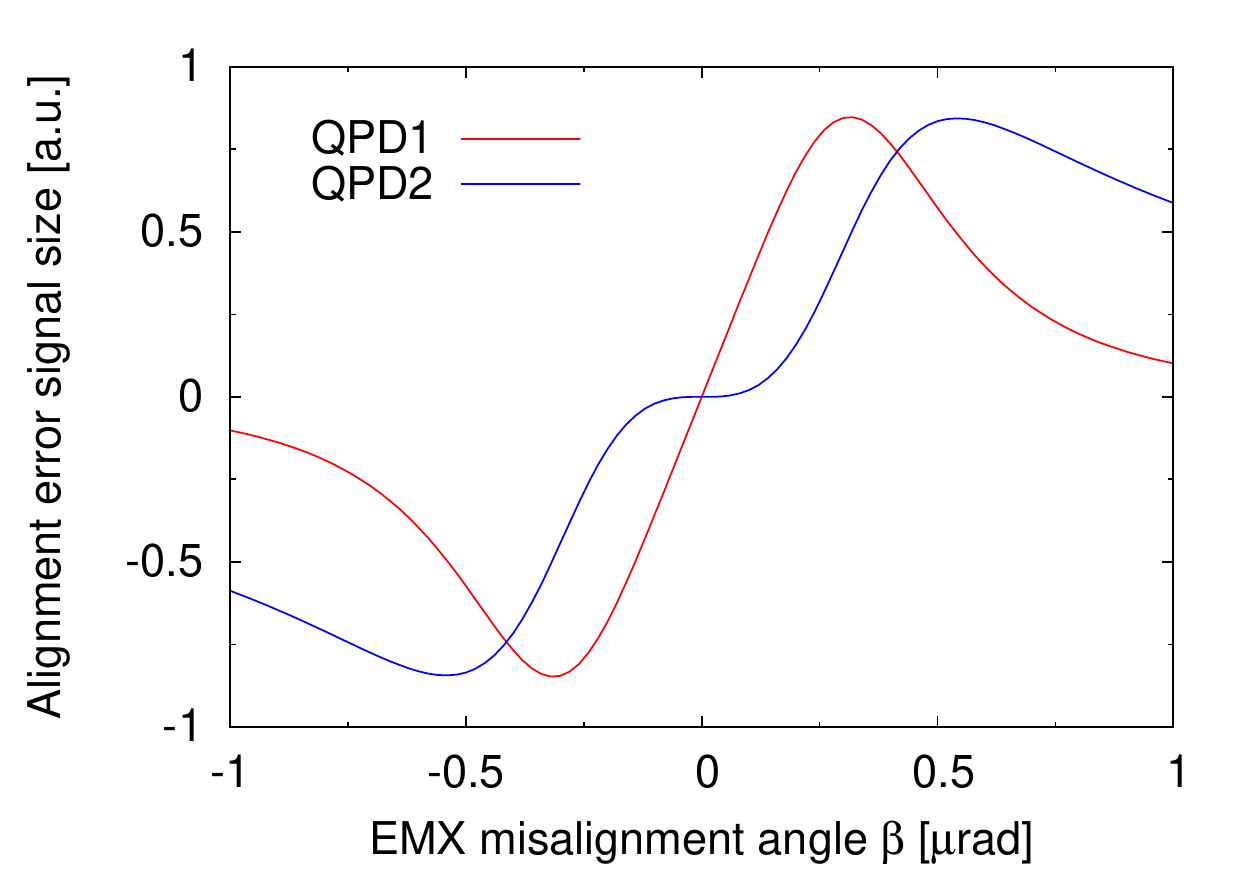}
\caption{Comparison of the obtained alignment error signals in the
$\mathrm{LG_{00}^{small}}$ case sensed with the two quadrant
photodiodes while mirror IMX (top) and mirror EMX (bottom) is
misaligned.} \label{LG00small-fig09-rot-4quad.eps}
\end{figure}

The proper sensing and control of the alignment degrees of freedom
of a gravitational wave detector is critical for its successful
operation. Hence, a comparison analysis of the alignment error
signals of the individual arm cavities for the three different
transversal mode configurations defined in
Section~\ref{sec:configurations-of-interest} is needed and
presented in this section. Full alignment control systems of
advanced GW detectors are very complex and depend on the details
of the detector design. The concept however, is firmly based on
the control of resonant cavities and we can use a single Fabry
Perot cavity to test whether LG modes are compatible with current
alignment control systems. The analysis uses an optical layout as
shown in
Figure~\ref{paper-cavity-alignment.eps}\,\textcircled{\scriptsize{\textbf{A}}}.
An electro-optic modulator (EOM) imprinting a phase modulation
with frequency $\Omega$ in combination with two quadrant
photodiodes is used to generate alignment error signals for the
two arm cavity mirrors using the Ward technique described in
\cite{Morrison1994}. Here each quadrant photodiode is responsible
for obtaining an alignment error signal of one arm cavity mirror.
In the following we only consider mirror rotations around vertical
axis. The results however remain applicable for the tilt degree of
freedom of each mirror. The optimisation of the error signals does
not use the theoretical optimal parameters but is done by tuning
the parameters; this reflects realistic experimental procedures.
The two quadrant photodiodes are placed such that their Gouy phase
difference is $90^\circ$ leaving the total Gouy phase arbitrary.
Each photodiode current is then demodulated at frequency $\Omega$.
The demodulation phase is chosen to maximise the error signal
slope for the corresponding mirror. An example of the
corresponding alignment error signals for the misalignment of both
arm cavity mirrors using the $\mathrm{LG_{00}^{small}}$
configuration can be seen in
Figure~\ref{LG00small-fig09-rot-4quad.eps}.
A good way of comparing different alignment error signals is
looking at the resulting control matrix~\cite{mantovani08,
phd.Mantovani}. In our example the control matrix contains the
values of the alignment error signal slopes $\sigma$ in the
working point generated by the two quadrant photodiodes QPD1 and
QPD2 for a misalignment of the arm cavity mirrors IMX and EMX, see
Figure~\ref{paper-cavity-alignment.eps}\,\textcircled{\scriptsize{\textbf{A}}}.
The subscript of $\sigma$ indicates the readout diode and the
superscript refers to the mirror of interest. Hence, the control
matrix is given in general by the following expression
\begin{equation}
C_\mathrm{configuration}= \left(
\begin{array}{cc}
\sigma^\mathrm{IMX}_\mathrm{QPD1} & \sigma^\mathrm{EMX}_\mathrm{QPD1} \\
\sigma^\mathrm{IMX}_\mathrm{QPD2} & \sigma^\mathrm{EMX}_\mathrm{QPD2} \\
\end{array}
\right)\,.
\end{equation}
Applying the optimisation procedure for the alignment signals
described earlier for all three configurations of interest,
results in the following three control matrices.
\begin{eqnarray}
\hspace{-1cm} C_\mathrm{LG_{00}^{small}}&=& \left(
\begin{array}{cc}
  5.6152 & 0.0477 \\
  2.1607 & 3.5878 \\
\end{array}
\right)\nonumber \\ &=& \,5.6152
\left(%
\begin{array}{cc}
  1 & 0.009 \\
  0.385 & 0.639 \\
\end{array}
\right)\\
\hspace{-1cm} C_\mathrm{LG_{33}}&=& \left(
\begin{array}{cc}
  7.444 & 0.022 \\
  2.741 & 4.771 \\
\end{array}%
\right) \nonumber \\ &=& \,7.444
\left(%
\begin{array}{cc}
  1 & 0.003 \\
  0.368 & 0.641 \\
\end{array}
\right)\\
\hspace{-1cm} C_\mathrm{LG_{00}^{large}}&=& \left(
\begin{array}{cc}
  17.774 & 15.330 \\
  11.472 & 2.725 \\
\end{array}%
\right) \nonumber \\ &=& \,17.774
\left(%
\begin{array}{cc}
  1 & 0.862 \\
  0.645 & 0.153 \\
\end{array}
\right)
\end{eqnarray}

For an ideal control matrix all matrix elements on the diagonal
would be one and the off-diagonal elements are zero. Comparing the
resulting control matrices, one can see that the
$\mathrm{LG_{00}^{large}}$ configuration performs worse than the
other two: both mirrors couple much more strongly into QPD1 than
into QPD2 making the alignment error signals far from ideal. The
$\mathrm{LG_{00}^{small}}$ and the $\mathrm{LG_{33}}$
configuration show a much better and almost even performance. This
is represented by the fact that in both of these configurations
the misalignment of mirror IMX couples three times more strongly
into QPD1 than into QPD2. Any misalignment of mirror EMX couples a
factor of 75 stronger into QPD2 compared to the signal sensed with
QPD1 for the $\mathrm{LG_{00}^{small}}$ configuration. This factor
increases further to 216 in the case of the $\mathrm{LG_{33}}$
configuration.

In conclusion we can say that the $\mathrm{LG_{00}^{large}}$ is
outperformed by the other two. The reason for this is not to be
found in the mode shape but in the cavity geometry. The RoCs of
the mirrors of the $\mathrm{LG_{00}^{small}}$ and the
$\mathrm{LG_{33}}$ configurations are the same, which results in
almost the same control matrix. In contrast the RoCs of the
mirrors of the $\mathrm{LG_{00}^{large}}$ configuration are much
smaller. For the $\mathrm{LG_{00}^{large}}$ configuration one
obtains a cavity g-parameter\cite{Siegman} of $g=0.91$ which is
very close to the instability border of unity. The other two
configurations have a g-factor of 0.33 which corresponds to a much
more stable and robust geometry.

\subsection{Coupling of differential arm cavity misalignment into the output port power}
\begin{figure}[t]
\centerline{
\includegraphics[width=8.6cm,keepaspectratio]{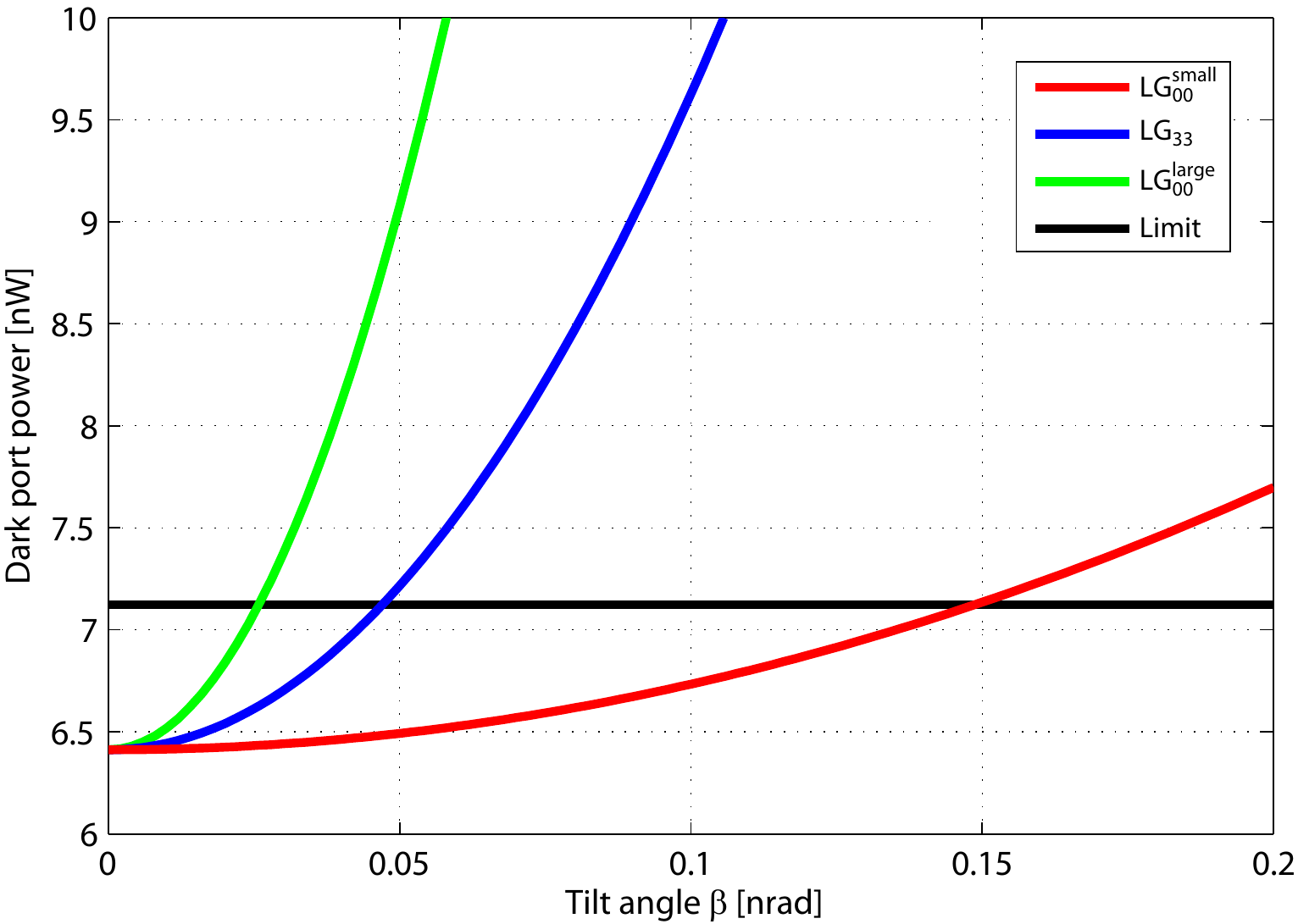}}
\caption{Comparison of the dark port power when the arm cavities
are differentially misaligned (see
Fig.~\ref{paper-cavity-alignment.eps}\,\textcircled{\scriptsize{\textbf{B}}})
in reference to the dark fringe power that results from a
differential arm length deviation of $10^{-15}$\,m while the
interferometers dark fringe offset is $10^{-12}$\,m.}
\label{fig6_HR_all_det.eps}
\end{figure}
An important measure for the performance of the different optical
modes in a full interferometer configuration is how much the
differential misalignment of the arm cavities and the
corresponding position change of each cavity's eigenmode couples
with the power in the interferometers output port due to the
reduce mode overlap on the beam splitter. A sketch of this
behaviour can be seen in
Figure~\ref{paper-cavity-alignment.eps}\,\textcircled{\scriptsize{\textbf{B}}}.
This coupling mechanism can generate a signal which is
indistinguishable from a GW signal. Hence it is important to
analyse how higher-order Laguerre Gauss modes compete with the
currently used fundamental mode. The following analysis compares
this coupling mechanism for the three configurations of interest
with respect to a reference value. This reference value is the
limit Advanced LIGO \cite{ADVLIGO} specifies for the differential
arm length deviation which is supposed to be smaller than
$10^{-15}$\,m \cite{LIGO-T070236-00-D}. This deviation together
with the envisaged dark fringe offset of $10^{-12}$\,m
\cite{LIGO-T070247-00-I} allows us to calculate the expected
increase in power in the output port of the interferometer to be
$7.124\cdot10^{-9}$\,W. This reference value can now be compared
to the output power enhancement of the three configurations of
interest for a misalignment of up to 200\,prad (see
Figure~\ref{fig6_HR_all_det.eps}). The $\mathrm{LG_{00}^{small}}$
configuration shows the smallest coupling from differential tilt
of the arm cavities into the output port power followed by the
$\mathrm{LG_{33}}$ and finally the $\mathrm{LG_{00}^{large}}$
configuration. The values where each configuration crosses the
reference limit of $7.124\cdot10^{-9}$\,W are summarised in
Table~\ref{tab:ifo-misalignment-to-power}.
\begin{table}[h!]
\centerline{
\begin{tabular}{l|c|c}
 & misalignment angle $\beta$ & coupling scaling factor\\
\hline \hline
$\mathrm{LG_{00}^{small}}$ & 148\,prad & 1\\
$\mathrm{LG_{33}}$ &  \phantom{1}46\,prad & 3.2\\
$\mathrm{LG_{00}^{large}}$ & \phantom{1}26\,prad & 5.7 \\
\end{tabular}
} \caption{Results of the coupling analysis between the
differential arm cavity misalignment and the interferometers
output port power. For each of the three mode configurations of
interest the misalignment angle is given at which the
interferometer output power has risen from it nominal value of
$6.412\cdot10^{-9}$\,W to the reference limit of
$7.124\cdot10^{-9}$\,W. The given coupling scaling factor shows
the relative coupling strength of each configuration referenced to
the $\mathrm{LG_{00}^{small}}$ configuration.}
\label{tab:ifo-misalignment-to-power}
\end{table}

In conclusion the $\mathrm{LG_{00}^{small}}$ configuration
performs at least a factor of 3.2 better than the other two
configurations. Nevertheless if we compare the two configurations
which use the same mirror size, we find that the
$\mathrm{LG_{33}}$ performes much better than the
$\mathrm{LG_{00}^{large}}$ configuration.

Overall we can conclude from the performed phase coupling
comparison analysis that higher-order Laguerre Gauss modes are
suitable for the implementation in future GW detectors. The direct
comparison between the two configurations which use the same
mirror sizes -- the $\mathrm{LG_{00}^{large}}$ and the
$\mathrm{LG_{33}}$ configuration -- clearly shows that in all of
the presented phase coupling analyses the configuration using
higher-order Laguerre Gauss modes performs better than the
configuration based on the traditionally used fundamental mode.
This serves to underline the great potential and prospects of
using higher-order Laguerre Gauss modes in future GW detectors.

\section{Prospects for Advanced Virgo}
\label{sec:prospects-Adv-Virgo}

In this section we focus on the currently planned second
generation GW detector Advanced Virgo \cite{ADVVIRGO} and its
sensitivity. The design efforts for Advanced Virgo are rapidly
progressing, yielding frequently improved detector configurations.
Hence the configuration presented in the following is unlikely to
be the final Advanced Virgo configuration, but rather represents a
snapshot of a development process. The numerical computations of
the detector sensitivity have been performed with
GWINC\footnote{In our analysis presented here we use a
specifically to Advanced Virgo adapted version 
of the simulation program GWINC -- Gravitational Wave
Interferometer Noise Calculator.} \cite{GWINC}.

In the following we compare the expected sensitivity for a
configuration using the fundamental TEM$_{00}$ mode against a
configuration using a LG$_{33}$ mode in three different scenarios
in which we assumed a fixed mirror radius of
$r_\mathrm{mirror}=0.17$\,m\footnote{In principle the Virgo mirror
radius is $r_\mathrm{mirror}=0.175$\,m, but a phase of 5\,mm
around the mirror limits the actual coating radius to
$\rho=0.17$\,m. Therefore the actual clipping loss of the
reflected beam has to be calculated using this coating radius.}.
The three scenarios are:
\begin{enumerate}
\item The beams of both mode configurations experience an
identical clipping loss at the ITM/ETM mirrors. Hence, the beam
size used in each configuration is different which results in a
different RoC of the arm cavity mirrors. (see
Sec.~\ref{sec:initial-considerations}). This means that for a
change of the mode shape used in the interferometer from e.g.
initially the TEM$_{00}$ mode to the LG$_{33}$ mode, all
interferometer mirrors have to be exchanged.

\item Both mode configurations use arm cavity mirrors with an
identical RoC. This enables a simple switching between the two
different mode shapes.

\item The planned thermal compensation system (TCS) is used to
change the RoC of the arm cavity mirrors by a certain amount.
Hence one can start with one mode configuration which uses arm
cavity mirrors with the optimal RoC. Later the TCS system enables
us to change the RoC to optimise the clipping loss of the other
mode configuration. Hence, this scenario is divided into two
parts. The first one starts with an optimised parameter set for
the TEM$_{00}$ mode and the TCS system is used to implement the
possible LG$_{33}$ mode configuration. In the second part of this
scenario we will start with an optimised parameter set for the
LG$_{33}$ mode and than change to a TEM$_{00}$ mode configuration
using the TCS system.
\end{enumerate}

The figures of merit of the comparison are the resulting effective
detection range for a binary neutron star inspiral
$\Gamma_\mathrm{NS/NS}$ and effective detection range for a binary
black hole inspiral $\Gamma_\mathrm{BH/BH}$.

\subsection{Scenario (i): Identical clipping loss}

The two mode configurations which are compared in this scenario
will be the reference configurations throughout the whole analysis
because of their fixed clipping loss of 1\,ppm. According to
Section~\ref{sec:initial-considerations} the beam sizes of each
configuration are different as well as the RoC of the
interferometer mirrors. This scenario and its two configurations
allow us to see the potential of using higher-order Laguerre Gauss
modes in future GW detectors for a fixed mirror size.

\begin{table*}[htbp]
    \begin{center}
    \begin{ruledtabular}
        \begin{tabular}{c||cccccc}
Scenario& SR det. & $w$ in  & $l_\mathrm{clip}$  & RoC & $\Gamma_\mathrm{NS/NS}$  & $\Gamma_\mathrm{BH/BH}$\\
(i) & [Hz] & [cm] & [ppm] & [m] & [Mpc] & [Mpc]\\
\hline \hline
TEM$_{00}$ & 750 & 6.47 & 1 & 1522.8 & 139.83 & 1135.2\\
LG$_{33}$  & 750 & 3.94 & 1 & 1708.4 & 168.34 & 1373.9\\
TEM$_{00}$ & 300 & 6.47 & 1 & 1522.8 & 148.85 & 1076.2\\
LG$_{33}$  & 300 & 3.94 & 1 & 1708.4 & 191.26 & 1322.4\\
        \end{tabular}
\caption{Input parameters and results of the GWINC simulation
analysis of the scenario (i) which uses identical clipping loss at
the arm cavity mirrors for each mode configuration.}
        \label{bench-results-scenario-one}
        \end{ruledtabular}
    \end{center}
\end{table*}

Table \ref{bench-results-scenario-one} comprises the input
parameters used in the analysis as well as the resulting inspiral
ranges. As one can see the analysis was performed for two
different signal recycling (SR) detunings \cite{Hild07a} -- 750Hz
and 300Hz -- to emphasise that the improvements vary with the
detuning. For both detunings the improvement using the LG$_{33}$
mode is significant. The total mirror thermal noise is decreased
by a
factor of up to 1.68. %
According to our 
simulation this results in a relative improvement of the inspiral
ranges by 20\%-21\% for a SR detuning of 750\,Hz and 23\%-28\% for
a SR detuning of 300\,Hz.
Hence the potential event rate\footnote{Assuming an uniform
distribution of sources in space, the event rate of a
gravitational detector is proportional to the volume of space it
can sense and therefore grows with the cube of the distance to a
detectable source.} of the Advanced Virgo detector can be
increased by a factor of up to 2.1 by using the LG$_{33}$ mode
instead of the fundamental TEM$_{00}$ mode.

\subsection{Scenario (ii): Identical RoCs}

\begin{table*}[htbp]
    \begin{center}
    \begin{ruledtabular}
        \begin{tabular}{c||cccccc}
Scenario& SR det. & $w$ in  & $l_\mathrm{clip}$  & $R_c$ & $\Gamma_\mathrm{NS/NS}$  & $\Gamma_\mathrm{BH/BH}$\\
(ii) & [Hz] & [cm] & [ppm] & [m] & [Mpc] & [Mpc]\\
\hline \hline
TEM$_{00}$ & 750 & 4.22 & 8.0e-9  & 1647.2 & 110.86 & 900.47\\
LG$_{33}$  & 750 & 4.22 & 30      & 1647.2 & 149.25 & 1375.3\\
        \end{tabular}
\caption{Input parameters and results of the GWINC simulation
analysis of the scenario (ii) which uses mirrors with an identical
RoC for each mode configuration.}
        \label{bench-results-scenario-two}
        \end{ruledtabular}
    \end{center}
\end{table*}

Our second scenario uses arm cavity mirrors with a fixed RoC
leading to the same beam size at the mirrors for both mode
configurations. The advantage of this configuration is that it
allows using either TEM$_{00}$ or the LG$_{33}$ mode without
exchanging the interferometer mirrors. To achieve this with a
reasonable sensitivity for both configurations, a tradeoff
concerning the clipping loss has to be made, see
Section~\ref{sec:initial-considerations} for more details. As a
result the beam size of the TEM$_{00}$ mode will be smaller
compared to the reference configuration. On the one hand this
results in much smaller clipping loss for this configurations, but
on the other hand the clipping loss of the LG$_{33}$ mode
configuration will go down to an acceptable value as well. The
input parameters and the results of the 
simulation analysis for this scenario are shown in
Table~\ref{bench-results-scenario-two}. The comparison of these
configurations shows that the inspiral ranges in the case of the
LG$_{33}$ mode configuration are greater by at least 35\% and up
to 53\%.
Although this looks promising, these two configurations have to be
compared with the reference configurations for the same SR
detuning analysed in scenario (i). We then find that the LG$_{33}$
configuration of scenario (ii) performs 0.1\% better concerning
the BH/BH inspiral range but the NS/NS inspiral range is 11\%
worse because of its higher clipping loss of
$l_\mathrm{clip}=30$\,ppm due to the slightly larger beam size.
This results in a decreased intra-cavity power which lowers the
sensitivity of the detector. The TEM$_{00}$ configuration of
scenario(ii) is also much less sensitive compared to the reference
configurations because of the small beam size and the resulting
higher thermal noise. The inspiral ranges of the TEM$_{00}$
configuration decrease by approximately 22\% in comparison to the
reference configuration.
The weak performance of the both configurations renders this
scenario not very favourable to be implemented into Advanced
Virgo.

\subsection{Scenario (iii): Use TCS to adapt RoCs}

Our third scenario combines the advantages of the two earlier
scenarios -- high performance and compatibility between the
different mode configurations. We propose to use the TCS, which is
an essential part of future GW detectors, to introduce a constant
offset onto the RoC of the arm cavity mirrors. This technique has
already been demonstrated in the GEO\,600 detector to match the
RoC of the two interferometer end mirrors \cite{Lueck04}. The
basic idea of this approach is to start with one mode
configuration with a clipping loss of $l_\mathrm{clip}=1$\,ppm at
each arm cavity mirror. The TCS will then be used later to change
the RoC of the arm cavity mirrors to optimise the useable beam
width for the corresponding other transversal mode. The currently
planned TCS \cite{LIGO-T080026-00-D} for second generation GW
detectors are designed to introduce a RoC change of $\Delta R_c=
-120$\,m. The TCS uses a ring heater placed near the mirror
substrate. The thermal radiation produced, is partly absorbed by
the mirror substrate thus deforming its original shape. By placing
the ring heater either in front or behind the mirror one can
actually change the sign of the RoC change allowing an adjustment
of $\Delta R_c\pm 120$\,m. The two different signs of the possible
RoC offset allow two different approaches whose input parameters
and results are comprised in
Table~\ref{bench-results-scenario-three}.

\begin{table*}[htbp]
    \begin{center}
    \begin{ruledtabular}
        \begin{tabular}{c||cccccc}
Scenario& SR det. & $w$ in  & $l_\mathrm{clip}$  & $R_c$ & $\Gamma_\mathrm{NS/NS}$  & $\Gamma_\mathrm{BH/BH}$\\
(iii) & [Hz] & [cm] & [ppm] & [m] & [Mpc] & [Mpc]\\
\hline \hline
TEM$_{00}$ & 750 & 6.47 & 1       & 1522.8 & 139.83 & 1135.2\\
LG$_{33}$  & 750 & 4.25 & 40.9    & 1642.2 & 142.93 & 1345.6\\
LG$_{33}$  & 750 & 3.94 & 1       & 1708.4 & 168.34 & 1373.9\\
TEM$_{00}$ & 750 & 4.71 & 4.8e-6  & 1588.2 & 118.40 & 960.98\\
        \end{tabular}
\caption{Input parameters and results of the GWINC simulation
analysis of the scenario (iii) which uses the same clipping loss
at the arm cavity mirrors for each mode configuration.}
        \label{bench-results-scenario-three}
        \end{ruledtabular}
    \end{center}
\end{table*}

The first approach uses the reference TEM$_{00}$ mode
configuration with $l_\mathrm{clip}=1$\,ppm as initial
interferometer configuration. This configuration with an arm
cavity RoC of $R_c=1522.8$\,m has already been analysed in
scenario (i). A constant change of the arm cavity RoC by $\Delta
R_c=+120$\,m introduced by the TCS will minimise the beam size of
a LG$_{33}$ mode. This minimised mode still experiences a clipping
loss of $l_\mathrm{clip}=40.9$\,ppm. Please note that in this case
the clipping losses of the LG$_{33}$ mode will go down further
with a larger RoC offset. The major advantage of this approach is
that we can use the reference TEM$_{00}$ mode configuration
initially and than later exchange it for a better performing
LG$_{33}$ mode configuration. This is reflected in the resulting
inspiral ranges of these two configurations (see
Table~\ref{bench-results-scenario-three}).
We find that the NS/NS and the BH/BH inspiral ranges of the
LG$_{33}$ mode configuration are increased by 2\% and 18\%,
compared to the reference TEM$_{00}$ mode configuration. For a
higher RoC change $\Delta R_c$ we can expect a much better
performance due to the lower clipping loss. To reach the reference
LG$_{33}$ mode configuration from scenario (i) which has a
clipping loss of $l_\mathrm{clip}=1$\,ppm, requires for example an
induced RoC change by the TCS of $\Delta R_c=186$\,m.

The second approach uses the TCS to start with the reference
LG$_{33}$ mode configuration with $l_\mathrm{clip}=1$\,ppm and an
arm cavity mirror RoC of $R_c=1708.4$\,m (see scenario (i)). The
TCS can be used to introduce a constant offset in the arm cavity
mirror RoC of $\Delta R_c=-120$\,m, which maximises the possible
beam size of a corresponding TEM$_{00}$ mode configuration to
$w=4.71$\,cm corresponding to the very small clipping loss
$l_\mathrm{clip}=$4.8e-6\,ppm. If the TCS is able to change the
RoC by a larger amount, the usable beam size of the TEM$_{00}$
mode configuration could be increased further.
This approach leads to an increase of the inspiral ranges of the
LG$_{33}$ mode configuration by between 42\% and 43\% compared to
the TEM$_{00}$ mode configuration. Despite these large
improvements with the LG$_{33}$ mode configuration it is
interesting to compare the TEM$_{00}$ mode configuration used here
to the reference TEM$_{00}$ mode configuration used in scenario
(i) to see how much one would lose by using this second approach
of scenario (iii). It turns out that the inspiral ranges go down
by approximately
15\%.

In conclusion scenario (iii) is the most promising one and the
idea to use the TCS to introduce a constant change to the RoC of
the arm cavity mirrors has great potential. It would allow a
change from a TEM$_{00}$ mode configuration to a LG$_{33}$ mode
configuration without exchanging the main optics of the
interferometer. To decide which of the two approaches described is
the better one, one has to judge if one could either afford to
have clipping losses for the LG$_{33}$ mode configuration of
$l_\mathrm{clip}=40.9$\,ppm in the first approach, or to have a
decreased performance of 15\% by the TEM$_{00}$ mode configuration
in the second approach compared to the reference TEM$_{00}$ mode
configuration.

\subsection{Sensitivity improvement from the use of LG$_{33}$ on the example
of Advanced Virgo }

\begin{figure}[t]
\centerline{
\includegraphics[width=8.6cm,keepaspectratio]{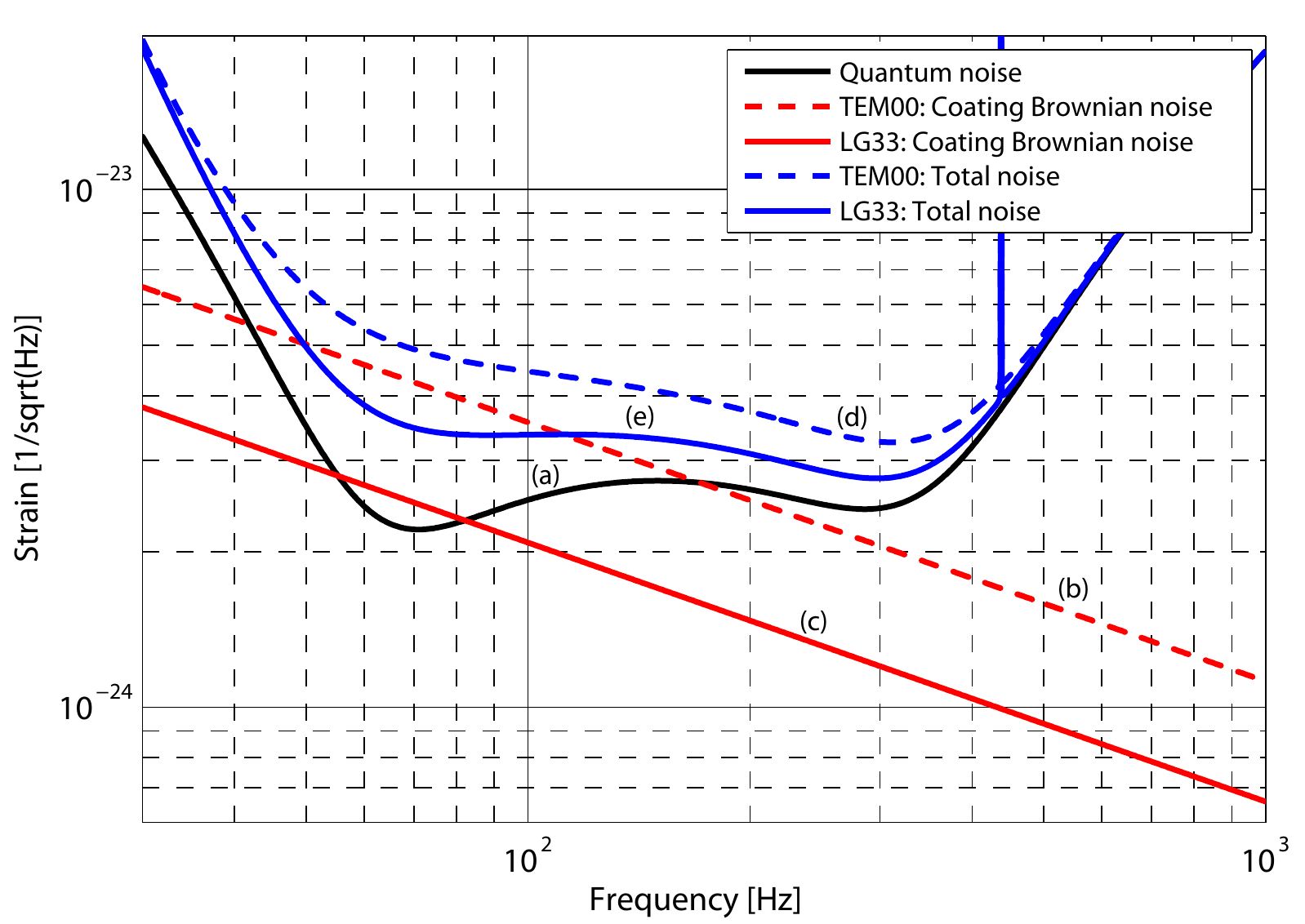}
} \caption{Sensitivity improvement from the implementation of the
LG$_{33}$ mode into the Advanced Virgo detector. This analysis is
based on the detector configuration described in the Advanced
Virgo Preliminary design \cite{VIR-089A-08}. For simplicity only
the contributions of coating Brownian noise (b, c) and quantum
noise (a) are shown, while all other noise contributions are
omitted in the plot, but taken into account for the overall
sensitivity (d, e). The sensitivity improvement when going from
TEM$_{00}$ (d) to LG$_{33}$ (e) mode corresponds to an improvement
of the binary neutron star inspiral range from 145 to 185\,Mpc and
increase the detectors NS/NS event rate by a factor of 2.1.}
\label{fig:Adv-now}
\end{figure}

As an illustrating example we evaluate how much the sensitivity of
an detector like Advanced Virgo could be improved by the
application of LG$_{33}$ modes. In particular we compare
configurations featuring TEM$_{00}$ and LG$_{33}$ modes with
identical clipping losses, but RoCs optimised for the individual
modes. Our analysis made use of the same parameters as the most
recent and comprehensive noise model for  Advanced Virgo
\cite{VIR-055A-08}, \cite{VIR-089A-08}. We assumed the following
reduction factors for the LG$_{33}$ mode in our GWINC simulation:
The coating Brownian and the substrate Brownian noise are reduced
by factors 1.7 and 1.9\footnote{These numbers can be calculated
using the formulas for the coating and substrate Brownian thermal
noise for finite sized mirrors \cite{Vinet09} in conjunction with
the following parameters. $w_\mathrm{LG00}=6.47$\,cm;
$w_\mathrm{LG33}=3.94$\,cm; $r_\mathrm{mirror}=0.17$\,m; mirror
thickness $h = 0.2$\,m; $\lambda=1064$\,nm; $f = 1$\,Hz; $T =
300$\,K; number of $\mathrm{SiO_2}$ coating layers
$N_\mathrm{SiO_2}=21$; number of $\mathrm{Ta_2O_5}$ coating layers
$N_\mathrm{Ta_2O_5}=20$; $\phi_\mathrm{substrate}=5e^{-9}$;
$\phi_\mathrm{SiO_2}=5e^{-5}$; $\phi_\mathrm{Ta_2O_5}=2e^{-4}$;
$\sigma_\mathrm{substrate}=0.167$;
$Y_\mathrm{substrate}=72.7$\,GPa; refractive indices
$n_\mathrm{SiO_2}=1.44876$ and $n_\mathrm{Ta_2O_5}=2.06$},
respectively, while the thermo-elastic noise increases by a factor
1.7 \cite{VINET}.

Figure \ref{fig:Adv-now} displays the resulting sensitivity curves
(blue traces) of Advanced Virgo featuring TEM$_{00}$ (dashed line)
and LG$_{33}$ (solid line) modes. As coating Brownian noise is
directly limiting the Advanced Virgo sensitivity in the frequency
range between 40 and 200\,Hz, an impressive sensitivity
improvement can be achieved by application of the LG$_{33}$ mode.
This broadband sensitivity improvement is concentrated on the
range from 30 to 400\,Hz with a maximal gain around 75\,Hz. The
binary neutron star inspiral
 range increases from 145 to 185\,Mpc, which corresponds to a rise of the expected
NS/NS event rate by a factor of 2.1.

\section{Summary and Outlook}
We carried out a comprehensive analysis of the prospects of
higher-order LG modes for future gravitational wave detectors.
Using numerical interferometer simulations, we compared the
behaviour of the LG$_{33}$ mode with the fundamental mode
(TEM$_{00}$). Our analysis included tilt to longitudinal phase
coupling, generation of angular control signals and the
corresponding control matrices for a single Fabry-Perot cavity as
well as the coupling of differential arm cavity misalignment into
dark port power for a full Michelson interferometer with arm
cavities. We were able to show that the LG$_{33}$ mode performs
similar if not even better than the commonly used TEM$_{00}$ for
all considered aspects of interferometric sensing. This strongly
indicates that all currently available experience and technology
for interferometric sensing and control, which is based on the
TEM$_{00}$ mode, can be transferred to the use of the LG$_{33}$
mode. Changing over from the fundamental mode to the LG$_{33}$
will not require any fundamental changes of the interferometric
control strategy or the control hardware, but only small
adjustments of the involved parameters, such as servo gains.

In addition, we performed a quantitative evaluation of the
expected sensitivity improvement from application of the LG$_{33}$
mode, using the planned Advanced Virgo detector as an example.
Three different options of how to change over from  the TEM$_{00}$
to the LG$_{33}$ mode have been developed. The first scenario
considers replacing the main mirrors, by ones with radii of
curvature optimised for the  LG$_{33}$ mode, while the second
scenario assumes that the same mirrors are used for both modes,
resulting in different clipping losses. In the third scenario the
Advanced Virgo thermal compensation system is used to adjust the
mirror curvatures for the two optical modes of interest
independently. The maximum sensitivity improvement is found to be
achievable when replacing the mirrors (scenario (i)). Using the
latest design parameters of Advanced Virgo we were able to show
that the application of the  LG$_{33}$ mode can give a broadband
improvement of the Advanced Virgo sensitivity for all frequencies
in the range from 30 to 400\,Hz. The corresponding binary neutron
star inspiral range increases from 145 to 185\,Mpc, enhancing the
potential detection rate of binary neutron star inspirals by a
factor 2.1.

The next steps towards the actual implementation of higher-order
LG modes in future GW detectors have to include the  demonstration
of efficient generation of high power LG beams, followed by
setting up of tabletop experiments for  experimental verification
of the simulations presented in the first half of this article.

\section{Acknowledgement}
We would like to thank J.~Y.~Vinet for fruitful discussions and
for providing us with the thermal noise suppression factors for
the higher-order LG modes. This work has been supported by the
Science and Technology Facilities Council (STFC) and the European
Gravitational Observatory (EGO). This document has been assigned
the LIGO Laboratory document number LIGO-P0900006-v1.

\section*{References}
\bibliographystyle{apsrev}
\bibliography{LG_new_archive}

\end{document}